\documentclass[copyright,creativecommons]{eptcs}
\providecommand{\event}{EXPRESS/SOS 2015}

\usepackage{amsfonts,amssymb,amsmath}
\usepackage{stmaryrd}
\usepackage{pgf}
\usepackage{tikz}
	\usetikzlibrary{trees,decorations,arrows,automata,positioning,plotmarks}
\usepackage{xspace}
\usepackage{extarrows}
\usepackage{color}
\usepackage{paralist}
\usepackage{hyperref}
\usepackage{breakurl}
\usepackage{ifthen}

\usepackage{qualityCriteria}

\title{Analysing and Comparing Encodability Criteria}
\def\titlerunning{Analysing and Comparing Encodability Criteria}
\author{Kirstin Peters\thanks{Supported by funding of the Excellence Initiative by the German Federal and State Governments (Institutional Strategy, measure `support the best').}
	\institute{TU Dresden, Germany}
	\and Rob van Glabbeek
	\institute{NICTA\thanks{NICTA is funded by the Australian Government through the Department of Communications and the Australian Research Council through the ICT Centre of Excellence Program.}, Sydney, Australia}
	\institute{Computer Science and Engineering, UNSW, Sydney, Australia}
}
\def\authorrunning{K. Peters \& R.\,J. van Glabbeek}

\begin{document}

\maketitle

\begin{abstract}
	Encodings or the proof of their absence are the main way to compare process calculi. To analyse the quality of encodings and to rule out trivial or meaningless encodings, they are augmented with quality criteria.
	There exists a bunch of different criteria and different variants of criteria in order to reason in different settings. This leads to incomparable results.
	Moreover it is not always clear whether the criteria used to obtain a result in a particular setting do indeed fit to this setting.
	We show how to formally reason about and compare encodability criteria by mapping them on requirements on a relation between source and target terms that is induced by the encoding function.
	In particular we analyse the common criteria \emph{full abstraction}, \emph{operational correspondence}, \emph{divergence reflection}, \emph{success sensitiveness}, and \emph{respect of barbs}; \eg we analyse the exact nature of the simulation relation (coupled simulation versus bisimulation) that is induced by different variants of operational correspondence.
	This way we reduce the problem of analysing or comparing encodability criteria to the better understood problem of comparing relations on processes.
\end{abstract}

\section{Introduction}

Encodings are used to compare process calculi and to reason about their expressive
power. Encodability criteria are conditions that limit the existence of encodings. Their main
purpose is to rule out trivial or meaningless encodings, but they can also be used to limit attention
to encodings that are of special interest in a particular domain or for a particular purpose.
These quality criteria are the main tool in \emph{separation results}, saying that one calculus is
not expressible in another one; here one has to show that no encoding meeting these criteria
exists. To obtain stronger separation results, care has to be taken in selecting quality criteria
that are not too restrictive.  For \emph{encodability results}, saying that one calculus is
expressible in another one, all one needs is an encoding, together with criteria testifying for the
quality of the encoding. Here it is important that the criteria are not too weak.

In the literature various different criteria and different variants of the same criteria are employed to achieve separation and encodability results \cite{vG94a,Nestmann00,nestmannPierce00,Palamidessi03,PalamidessiSVV06, Nestmann06,CCAV08,Parrow08,BusiGZ09,Gorla10a,PN12,vG12}.
Some criteria, like full abstraction or operational correspondence, are used frequently.
Other criteria are used to enforce a property of encodings that might only be necessary within a certain domain.
For instance, the homomorphic translation of the parallel operator---in general a rather strict criterion---was used in \cite{Palamidessi03} to show the absence of an encoding from the synchronous into the asynchronous $ \pi $-calculus, because this requirement forbids for the introduction of global coordinators. Thus this criterion is useful when reasoning about the concurrent behaviour of processes, although it is in general too strict to reason about their interleaving behaviour. Unfortunately it is not always obvious or clear whether the criteria used to obtain a result in a particular setting do indeed fit to this setting. Indeed, as discussed in \cite{PN12}, the homomorphic translation of the parallel operator forbids more than global coordinators, \ie is too strict even in a concurrent setting.

The different purposes of encodability criteria lead to very different kinds of conditions that are usually hard to analyse and compare directly. In fact even widely used criteria---as full abstraction---seem not to be fully understood by the community, as the need for articles as \cite{gorla2014abstraction,parrow2014abstraction} shows. In contrast to that, relations on processes---such as simulations and bisimulations---are a very well studied and understood topic (see for example \cite{vanglabbeek01}). Moreover it is natural to describe the behaviour of terms, or compare them, modulo some equivalence relation. Also many encodability criteria, like operational correspondence, are obviously designed with a particular kind of relation between processes in mind. Therefore, in order to be able to formally reason about encodability criteria, to completely capture and describe their semantic effect, and to analyse side conditions of combinations of criteria, we map them on conditions on relations between source and target terms.

We consider the disjoint union $ \procS \uplus \procT $ of the terms or processes from the source and target languages of an encoding. Then we describe the effect an encodability criterion {\bf C} has on the class of permitted encoding functions in terms of a relation $ \IRel $ that relates at least all source terms to their literal translations, \ie contains the pair $ \left( S, \Encoding{S} \right) $ for all source terms $ S $. If the encodability criterion {\bf C} is defined \wrt some additional relations on the source or target languages, as it is the case for full abstraction and operational correspondence, we usually also include these relations in $ \IRel $. In order to completely capture the effect of a criterion {\bf C} we aim at iff-results of the form
\begin{quote}
	$ \encoding $ satisfies {\bf C} iff there exists a relation $ \IRel $ such that $ \forall S\logdot \left( S, \Encoding{S} \right) \in \IRel $ and $ \Pred{\IRel} $,
\end{quote}
where $ \pred $ is the condition that captures the effect of {\bf C}.
For example, an encoding reflects divergence iff there exists a relation $ \IRel $ such that $ \forall S\logdot \left( S, \Encoding{S} \right) \in \IRel $ and $ \IRel $ reflects divergence.

We illustrate this approach by applying it to some very common criteria. We start with divergence reflection in \S\ref{sec:divergence}, because it is simple and well understood. Accordingly, we do not gain significant new insights, but it suits us very well to introduce our approach. In the same way success sensitiveness and respect of barbs are analysed. We then switch to the criteria full abstraction in \S\ref{sec:fullAbstraction} and operational correspondence in \S\ref{sec:operationalCorrespondence}, which are possibly not completely understood yet. In particular, we show a connection between full abstraction and transitivity, and prove to which kinds of simulation relations common variants of operational correspondence are linked.
In \S\ref{sec:combining} we analyse the effects of combining the above criteria. Since we first map the criteria to conditions on relations between source and target terms, analysing their combined effect requires us to identify a suitable witness relation for the combined conditions. Combining divergence reflection and success sensitiveness is simple, as
illustrated in \S\ref{sec:divergenceAndSuccess}. Combining these two
criteria with operational correspondence (\S\ref{sec:addingOperationalCorrespondence}) is more elaborate.
Finally we analyse the effect of combining full abstraction with operational correspondence in \S\ref{sec:fullAbstractionAndOperationalCorrespondence}.

All claims in this paper have been proved using the interactive theorem prover Isabelle/HOL \cite{isabelle02}.
The Isabelle implementation of the theories is available in the `Archive of Formal Proofs' at\\[1ex]\mbox{}\hfill
\url{http://afp.sourceforge.net/entries/Encodability_Process_Calculi.shtml}.\hfill\mbox{}

\section{Technical Preliminaries}

We analyse criteria used to reason about the quality of encodings between process calculi. We do not force any limitations on the considered calculi. A \emph{process calculus} is a language $ \lang_C = \Tuple{\proc_C, \step_C} $ consisting of a set of terms $ \proc_C $---its \emph{syntax}---and a relation on terms $ \step_C \; \subseteq \proc_C \times \proc_C $---its \emph{semantics}.
The elements of $ \proc_C $ are called \emph{process terms} or shortly processes or terms.

Here we assume that the semantics of the language is provided as a so-called reduction semantics,
because in the context of encodings
the treatment of reductions is simpler---the consideration of labelled semantics and of criteria using labelled steps is left for further work.
A \emph{step} $ P \step_C P' $ is an element $ (P, P') \in \; \step_C $. Let $ \Steps[]{C} $ denote the reflexive and transitive closure of $ \step_C $. We write $ P \step_C $ if $ \exists P' \logdot P \step_C P' $ and $ P \StepsW{C} $ if $ P $ can do an infinite sequence of steps.
A term $ P $ such that $ P \StepsW{C} $ is called \emph{divergent}.

Languages can be augmented with (a set of) relations $ \Rel_C \subseteq \proc_C^2 $ on their processes.
If $ \Rel \subseteq B^2 $ is a relation and $ B' \subseteq B $, then $ \Red{\Rel}{B'} = \Set{ \left( x, y \right) \mid x, y \in B' \wedge \left( x, y \right) \in \Rel } $ denotes the restriction of $ \Rel $ to the domain $ B' $. A relation $ \Rel $ \emph{preserves} some
condition $ \pred: B \to \bool $ (with $\bool$ representing the Booleans) if whenever $ (P, Q) \in \Rel $ and $ P $ satisfies $ \pred $ then $ Q $ satisfies $ \pred $. A relation $ \Rel $
\emph{reflects} $ \pred $ if whenever $ (P, Q) \in \Rel $ and $ Q $ satisfies $ \pred $ then also $ P $ satisfies $ \pred $. Finally $ \Rel $ \emph{respects} a
condition $ \pred $ if $ \Rel $ preserves and reflects it. We use $ \refl{\cdot} $, $ \sym{\cdot} $, and $ \trans{\cdot} $ to denote the \emph{reflexive}, \emph{symmetric}, and
\emph{transitive closure} of a binary relation, respectively.

Relations on process terms are an important tool to reason about processes and languages. Of special interest are simulation relations; in particular bisimulations.
$ \Rel $ is a bisimulation if any two related processes mutually simulate their respective sequences of steps, such that the derivatives are again
related.

\begin{definition}[Bisimulation]
	$ \Rel $ is a \emph{(weak reduction) bisimulation} if for each $ (P, Q) \in \Rel $:
	\begin{compactitem}
		\item $ P \steps P' $ implies $ \exists Q' \logdot Q \steps Q' \wedge (P', Q') \in \Rel $
		\item $ Q \steps Q' $ implies $ \exists P' \logdot P \steps P' \wedge (P', Q') \in \Rel $
	\end{compactitem}
	Two terms are \emph{bisimilar} if there exists a bisimulation that relates them.
\end{definition}

\noindent
The definition of a \emph{strong (reduction) bisimulation} is obtained by replacing all $ \steps $ by $ \step $ in the above definition, \ie a strong bisimulation requires that a step has to be simulated by a single step.
Coupled similarity is strictly weaker than bisimilarity. As pointed out in \cite{parrowCoupled92}, in contrast to bisimilarity it allows for intermediate states in simulations: states that cannot be identified with states of the simulated term. Each symmetric coupled simulation is a bisimulation.

\begin{definition}[Coupled Simulation]
	A relation $ \Rel $ is a \emph{(weak reduction) coupled simulation} if both $ \left( \exists Q' \logdot Q \steps Q' \wedge (P', Q') \in \Rel \right) $ and $ \left( \exists Q' \logdot Q \steps Q' \wedge (Q', P') \in \Rel \right) $ whenever $ (P, Q) \in \Rel $ and $ P \steps P' $.\\
	Two terms are \emph{coupled similar} if they are related by a coupled simulation in both directions.
\end{definition}

An \emph{encoding} from $ \langS = \Tuple{\procS, \stepS} $ into $ \langT = \Tuple{\procT, \stepT} $ relates two process calculi. We call $ \langS $ the \emph{source} and $ \langT $ the \emph{target language}. Accordingly, terms of $ \procS $ are \emph{source terms} and of $ \procT $ \emph{target terms}. In the simplest case an encoding from $ \langS $ into $ \langT $ is an \emph{encoding function} $ \encoding : \procS \to \procT $ from source terms into target terms.
Sometimes an encoding is defined by several functions, such as the encoding function and the renaming policy used in the framework of \cite{Gorla10a}.
Else we identify an encoding with its encoding function.

An \emph{encodability criterion} is a predicate on encoding functions, used to reason about the quality of encodings. We analyse such criteria by mapping them on requirements on relations $ \IRel \subseteq \left( \procS \uplus \procT \right)^2 $ on the disjoint union of the source and target terms of the considered encodings~$ \encoding $. To simplify the presentation we assume henceforth that $ \procS \cap \procT = \emptyset $ and thus $ \procS \uplus \procT = \procS \cup \procT $. The Isabelle proofs do not rely on such an assumption.
We say that a condition $ \pred : \left( \procS \uplus \procT \right) \to \bool $ is \emph{preserved} by an encoding if for all source terms $ S $
that satisfy $ \pred $, the condition $ \pred $ also holds for $ \Encoding{S} $. A condition is \emph{reflected} by an encoding if whenever $ \Encoding{S} $ satisfies it, then so does $ S $.
Finally an encoding \emph{respects} a condition if it both preserves and reflects it.

\section{Analysing Encodability Criteria}\label{analysing}
\label{sec:analysing}

An encoding function $ \encoding : \procS \to \procT $ maps source terms on target terms.
Thereby it induces a relation on the combined domain of source and target terms that relates source terms with their literal translations. We start with this relation, \ie in the simplest case we map an encodability criterion to a requirement on a relation $ \IRel \subseteq \left( \procS \uplus \procT \right)^2 $ that contains at least the pairs $ \left( S, \Encoding{S} \right) $ for all source terms $ S \in \procS $. If we consider a criterion that is defined \wrt some relations on the source or target, we will also include these relations in $ \IRel $, possibly closing the latter under reflexivity, symmetry, and/or transitivity.

Alternatively, we could require that $ \IRel $ relates source terms and their literal translations
in both directions, meaning that $ \left( S, \Encoding{S} \right) \in \IRel $ and $ \left( \Encoding{S},
S \right) \in \IRel $ for all source terms $ S \in \procS $. However, this condition limits our
analysis to properties that are respected. It does not allow us to reason about properties like
divergence reflection, where some condition need only to be reflected but not necessarily be
preserved, or vice versa. Accordingly we follow the first approach.

\subsection{Divergence Reflection and Observables}
\label{sec:divergence}

We start with divergence reflection as defined in \cite{Gorla10a}, because it is often easy to establish and well understood. An encoding reflects divergence if it does not introduce divergence, \ie if all divergent translations result from divergent source terms.

\begin{definition}[Divergence Reflection]
	An encoding $ \encoding : \procS \to \procT $ \emph{reflects divergence} if $ \Encoding{S} \stepsTW $ implies $ S \; \stepsSW $ for all source terms $ S \in \procS $.
\end{definition}

We can reformulate this criterion as follows: An encoding reflects divergence if it reflects the predicate $ \lambda P \logdot P \stepsW $. To analyse this criterion it suffices to consider the relation $ \Set{ \left( S, \Encoding{S} \right) \mid S \in \procS } $. It is obvious that an encoding reflects divergence iff $ \Set{ \left( S, \Encoding{S} \right) \mid S \in \procS } $ reflects divergence, \ie reflects the predicate $ \lambda P \logdot P \stepsW $. In fact we can generalise this case. If an encodability criterion can be described by the preservation or reflection of a predicate, then an encoding satisfies this criterion iff $ \Set{ \left( S, \Encoding{S} \right) \mid S \in \procS } $ preserves or reflects this predicate.
Of course direction ``if'' holds for any relation that contains at least the pairs $ \left( S, \Encoding{S} \right) $.
We use the relation $ \Set{ \left( S, \Encoding{S} \right) \mid S \in \procS } $ as a witness and it allows us to analyse the combination of different criteria later.

\begin{lemma}[Preservation]
	Let $ \pred : \left( \procS \uplus \procT \right) \to \bool $ be a predicate.
	An encoding preserves the predicate $ \pred $ iff $ \exists \IRel \logdot \left( \forall S \logdot \left( S, \Encoding{S} \right) \in \IRel \right) \wedge \IRel $ preserves $ \pred $.
\end{lemma}

\noindent
We obtain a similar result if we replace the unary predicate $ \Pred{\cdot} $ by the binary predicate $ \Pred{\cdot, \cdot} $ of type $ \left( \procS \uplus \procT \right) \times \mathcal{T} \to \bool $ for some arbitrary type $ \mathcal{T} $ to represent predicates with several parameters. Moreover we obtain the same result for either reflection or respect instead of preservation.

Accordingly an encoding reflects divergence, \ie the predicate $ \lambda P \logdot P \stepsW $, iff there exists a relation $ \IRel $ that relates at least each source term to its literal translation and reflects this predicate.

\begin{lemma}[Divergence Reflection]\label{lem:divergence}
	An encoding $ \encoding : \procS \to \procT $ from $ \langS $ into $ \langT $ reflects divergence iff  $ \exists \IRel \logdot \left( \forall S \logdot \left( S, \Encoding{S} \right) \in \IRel \right) \wedge \IRel $ reflects divergence.
\end{lemma}

In a similar way we can deal with the criterion barb sensitiveness.  A \emph{barb} is a property
of a process that is treated as an observable, and whose reachability should be respected by an encoding.
We assume that $ \mathcal{B} $ is a set of barbs that contains at least all barbs of the source and
the target language. Moreover we assume that each language $ \lang $ specifies its own predicate $
\HasBarbL{\lang}{\cdot}{\cdot} $ such that $ \HasBarbL{\lang}{P}{a} $ returns true if $ P \in
\proc_{\lang} $ and $ P $ has the barb $ a $ in $ \lang $. If a barb $ a $ is not relevant or
present in a language $ \lang $ then $ \HasBarbL{\lang}{P}{a} $ does not hold for any $ P \in
\proc_\lang $. We use $ \ReachBarbL{\lang}{P}{a} $ if $ P $ reaches the barb $ a $ in $ \lang $, \ie
$ \ReachBarbL{\lang}{P}{a} \deff \exists P'. P \steps_{\lang} P' \wedge \HasBarbL{\lang}{P'}{a} $.

An encoding weakly respects source term barbs iff it respects the predicate $ \lambda P \; a \logdot \ReachBarb{P}{a} $.
This holds iff $ \Set{ \left( S, \Encoding{S} \right) \mid S \in \procS } $
respects this predicate, which in turn is the case iff there exists a relation $ \IRel $ that
relates at least each source term to its literal translation and respects this predicate.

\begin{lemma}[Barb Sensitiveness]
	Assume $ \langS $ and $ \langT $ each define a predicate $ \HasBarb{\cdot}{\cdot} : \proc \times \mathcal{B} \to \bool $.\linebreak
	$ \encoding: \procS \to \procT $ weakly respects barbs iff $ \exists \IRel \logdot \left( \forall S \logdot \left( S, \Encoding{S} \right) \in \IRel \right) \wedge \IRel $ weakly respects barbs.
\end{lemma}

\noindent
Again we obtain a similar result if we replace respect by preservation or reflection or if we consider the existence instead of the reachability of barbs.

However, only very few encodings directly preserve or reflect barbs. More often barbs are
translated, as for example in the encodings between different variants of the $ \pi $-calculus
in \cite{nestmannPierce00, peters12} or the two translations from CSP into variants of CCS with name passing in \cite{cspToCcs15}. Since we do not fix the definition of $
\HasBarbL{\lang}{\cdot}{\cdot} $, this can for instance be expressed by adapting this predicate in the target language.

In a similar way we can deal with the criterion success sensitiveness. This criterion was proposed by Gorla as part of his encodability framework \cite{Gorla10a}. An encoding is success sensitive if it respects reachability of a particular process $ \success $ that represents successful termination, or some other form of success, and is added to the syntax of the source as well as the target language. We write $ P\hasSuccess $ to denote the fact that $ P $ is successful---however this predicate might be defined in the particular source or target language. Reachability of success is then defined as $ P\reachSuccess \deff \exists P' \logdot P \steps P' \wedge P'\hasSuccess $. An encoding is success sensitive if each source term and its translation answer the test for reachability of success in the same way.

\begin{definition}[Success Sensitiveness]
	Let $ \langS $ and $ \langT $ each define a predicate $ \cdot\hasSuccess : \proc \to \bool $.
	An encoding $ \encoding : \procS \to \procT $ is \emph{success sensitive} if, for all $ S \in \procS $, $ S\reachSuccess $ iff $ \Encoding{S}\reachSuccess $.
\end{definition}

\noindent
Accordingly, an encoding is success sensitive iff it respects the predicate $ \lambda P \logdot P\reachSuccess $. This is the case iff $ \Set{ \left( S, \Encoding{S} \right) \mid S \in \procS } $ respects this predicate, which in turn is the case iff there exists a relation $ \IRel $ that relates at least each source term to its literal translation and respects this predicate.

\begin{lemma}[Success Sensitiveness]\label{lem:success}
	Assume $ \langS $ and $ \langT $ each define a predicate $ \cdot\hasSuccess : \proc \to \bool $.
	An encoding $ \encoding : \procS \to \procT $ is success sensitive iff $ \exists \IRel \logdot \left( \forall S \logdot \left( S, \Encoding{S} \right) \in \IRel \right) \wedge \IRel $ respects $ \lambda P \logdot P\reachSuccess $.
\end{lemma}

Success sensitiveness links source term behaviours to behaviours of target terms. If the source and
the target language are very different, they can impose quite different kinds of behaviour that
might be hard to compare directly. For example, observables in the $ \pi
$-calculus refer to the existence of unguarded input or output prefixes \cite{milnerParrowWalker92},
whereas in the core of mobile ambients there are no in- or outputs but only ambients and action
prefixes that describe the entering, leaving, and opening of an ambient
\cite{cardelliGordon00}. Success sensitiveness allows to compare such languages by introducing
a new kind of barb that can be understood in both calculi.
If we want to compare two languages that are very similar, such as two variants of the same
calculus, we can demand stricter encodability criteria and compare their barbs directly.

Next we concentrate on criteria that cannot be expressed simply by the preservation or reflection of some predicate.

\subsection{Full Abstraction}
\label{sec:fullAbstraction}

Full abstraction was probably the first criterion that was widely used to reason about the quality of encodings \cite{riecke91, mitchell93, perez09}. This criterion is defined \wrt a relation $ \RelS \subseteq \procS^2 $ on source terms
and a relation 	$ \RelT \subseteq \procT^2 $ on target terms. An encoding is fully abstract \wrt $
\RelS $ and $ \RelT $ if two source terms are related by $ \RelS $ iff their literal translations
are related by $ \RelT $.

\begin{definition}[Full Abstraction]
	An encoding $ \encoding : \procS \to \procT $ is \emph{fully abstract} \wrt the relations $ \RelS \subseteq \procS^2 $ and $ \RelT \subseteq \procT^2 $ if, for all $ S_1, S_2 \in \procS $, $ \left( S_1, S_2 \right) \in \RelS $ iff $ \left( \Encoding{S_1}, \Encoding{S_2} \right) \in \RelT $.
\end{definition}

\noindent
There are a number of trivial full abstraction results, \ie results that hold for all (or nearly all) encodings (see \eg \cite{gorla2014abstraction,parrow2014abstraction}). In particular, for each encoding and each target term relation $ \RelT \subseteq \procT^2 $ there exits a source term relation $ \RelS \subseteq \procS^2 $, namely $ \Set{ (S_1, S_2) \mid (\Encoding{S_1}, \Encoding{S_2}) \in \RelT } $, such that the encoding is fully abstract \wrt $ \RelS $ and $ \RelT $. For each injective encoding and each source term relation $ \RelS \subseteq \procS^2 $, there exits $ \RelT \subseteq \procT^2 $, namely $ \Set{ (\Encoding{S_1}, \Encoding{S_2}) \mid (S_1, S_2) \in \RelS } $, such that the encoding is fully abstract \wrt $ \RelS $ and $ \RelT $. Accordingly we consider full abstraction \wrt fixed source and target term relations.

As suggested above, we map this criterion on a relation that relates at least each source term to its literal translation and includes the relations $ \RelS $ and $ \RelT $. If we additionally add pairs of the form $ \left( \Encoding{S}, S \right) $ for all $ S \in \procS $, we make an interesting observation. If we surround the pair $ \left( S_1, S_2 \right) \in \RelS $ by the pairs $ \left( \Encoding{S_1}, S_1 \right) $ and $ \left( S_2, \Encoding{S_2} \right) $ and add transitivity we obtain the pair $ \left( \Encoding{S_1}, \Encoding{S_2} \right) $. Similarly, from transitivity, $ \left( S_1, \Encoding{S_1} \right) $, $ \left( \Encoding{S_1}, \Encoding{S_2} \right) $, and $ \left( \Encoding{S_2}, S_2 \right) $ we obtain the pair $ \left( S_1, S_2 \right) $.
Because of this, an encoding is fully abstract \wrt the preorders $ \RelS $ and $ \RelT $ iff there exists a transitive relation $ \IRel $ that relates at least each source term to its literal translation in both directions, such that the restriction of $ \IRel $ to source/target terms is $ \RelS $/$ \RelT $.

\begin{lemma}[Full Abstraction]
	$ \encoding : \procS \to \procT $ is \emph{fully abstract} \wrt the preorders $ \RelS \subseteq \procS^2 $ and $ \RelT \subseteq \procT^2 $ iff $ \exists \IRel \logdot \left( \forall S \logdot \left( S, \Encoding{S} \right), \left( \Encoding{S}, S \right) \in \IRel \right) \wedge \RelS = \Red{\IRel}{\procS} \wedge \RelT = \Red{\IRel}{\procT} \wedge \IRel $ is transitive.
\end{lemma}

\noindent
Thus an encoding is fully abstract \wrt $ \RelS $ and $ \RelT $ if the encoding function combines the relations $ \RelS $ and $ \RelT $ in a transitive way.

In order to allow combinations with criteria like divergence reflection, \ie predicates that are not respected but preserved or reflected, we get rid of the requirement on the pairs $ \left( \Encoding{S}, S \right) $. Therefore we consider the symmetric closure of $ \IRel $.
An encoding is fully abstract \wrt the equivalences $ \RelS $ and $ \RelT $ iff there exists a
relation $ \IRel $ that relates at least each source term to its literal translation, such that
the restriction of the symmetric closure of $ \IRel $ to source/target terms is $ \RelS $/$ \RelT $
and the symmetric closure of $ \IRel $ is a preorder.

\begin{lemma}[Full Abstraction]
	An encoding $ \encoding : \procS \to \procT $ is fully abstract \wrt the equivalences $ \RelS \subseteq \procS^2 $ and $ \RelT \subseteq \procT^2 $ iff $ \exists \IRel \logdot \left( \forall S \logdot \left( S, \Encoding{S} \right) \in \IRel \right) \wedge \RelS = \Red{\sym{\IRel}}{\procS} \wedge \RelT = \Red{\sym{\IRel}}{\procT} \wedge \sym{\IRel} $ is a preorder.
\end{lemma}

\noindent
Since it is always possible to construct a relation that includes $ \RelS $, $ \RelT $, and pairs $
\left( S, \Encoding{S} \right) $, the crucial requirement on the right-hand side is transitivity.
A discussion of this criterion and references to earlier such discussions can be found in \cite{perez09, gorla2014abstraction}.

\subsection{Operational Correspondence}
\label{sec:operationalCorrespondence}

To strengthen full abstraction it is often combined with operational correspondence. This criterion requires that source terms and their translations `behave' similar, by requiring that steps are
preserved and reflected modulo some target term relation $ \RelT \subseteq \procT^2 $. Intuitively an encoding is operational corresponding \wrt $ \RelT $ if each source term step is simulated by its
translation, \ie $ \encoding $ does not remove source behaviour (\emph{completeness}), and each step of the target is part of the simulation of a source term step, \ie $ \encoding $ does not introduce
new behaviour (\emph{soundness}).
There are a number of different variants of this criterion. We consider three unlabelled variants \cite{nestmannPierce00,Gorla10a}. In particular the last variant, proposed in \cite{Gorla10a}, was used for numerous encodability and separation results.

\begin{definition}[Operational Correspondence]
	An encoding $ \encoding: \procS \to \procT $ is \emph{strongly operationally corresponding} \wrt $ \RelT \subseteq \procT^2 $ if it is:
	\begin{compactitem}
		\item[\; Strongly Complete:] $ \forall S, S' \logdot S \stepS S' \text{ implies } \left( \exists T \logdot \Encoding{S} \stepT T \wedge \left( \Encoding{S'}, T \right) \in \RelT \right) $
		\item[\; Strongly Sound:] $ \forall S, T \logdot \Encoding{S} \stepT T \text{ implies } \left( \exists S' \logdot S \stepS S' \wedge \left( \Encoding{S'}, T \right) \in \RelT \right) $
	\end{compactitem}
	$ \encoding: \procS \to \procT $ is \emph{operationally corresponding} \wrt $ \RelT \subseteq \procT^2 $ if it is:
	\begin{compactitem}
		\item[\; Complete:] $ \forall S, S' \logdot S \stepsS S' \text{ implies } \left( \exists T \logdot \Encoding{S} \stepsT T \wedge \left( \Encoding{S'}, T \right) \in \RelT \right) $
		\item[\; Sound:] $ \forall S, T \logdot \Encoding{S} \stepsT T \text{ implies } \left( \exists S' \logdot S \stepsS S' \wedge \left( \Encoding{S'}, T \right) \in \RelT \right) $
	\end{compactitem}
        $ \encoding: \procS \to \procT $ is \emph{weakly operationally corresponding} \wrt $ \RelT \subseteq \procT^2 $ if it is:
	\begin{compactitem}
		\item[\; Complete:] $ \forall S, S' \logdot S \stepsS S' \text{ implies } \left( \exists T \logdot \Encoding{S} \stepsT T \wedge \left( \Encoding{S'}, T \right) \in \RelT \right) $
		\item[\; Weakly Sound:] $ \forall S, T \logdot \Encoding{S} \stepsT T \text{ implies } \left( \exists S', T' \logdot S \stepsS S' \wedge T \stepsT T' \wedge \left( \Encoding{S'}, T' \right) \in \RelT \right) $
	\end{compactitem}
\end{definition}

\noindent
Again this criterion is trivial if we do not fix the target term relation. Each encoding is operational corresponding \wrt the universal relation on target terms.

The formulation of operational correspondence (in all its variants) strongly reminds us of simulation relations on processes, such as bisimilarity. Obviously this criterion is designed in order to establish a simulation-like relation between source and target terms. We now determine the exact nature of this relation.
The first two variants exactly describe strong and weak bisimilarity up to $ \RelT $. More precisely, an encoding is operational
corresponding \wrt a preorder $ \RelT $ that is a bisimulation iff there exists a preorder $ \IRel $, such as $ \trans{\refl{\Set{\left( S, \Encoding{S} \right) \mid S \in \procS} \cup \RelT}} $, that is a bisimulation, relates at least all source terms to their literal translations, and such that $ \RelT = \Red{\IRel}{\procT} $, and for all pairs $ (S, T) \in \IRel $ it holds that $ \left( \Encoding{S}, T \right) \in \RelT $. The last condition is necessary to ensure operational correspondence, and $ \RelT = \Red{\IRel}{\procT} $ ensures that $ \RelT $ is a bisimulation if $ \IRel $ is.
Accordingly, operational correspondence ensures that source terms and their translations are bisimilar.

\begin{lemma}[Operational Correspondence]
	\label{lem:operationalCorrespondence}
	An encoding $ \encoding : \procS \to \procT $ is operational corresponding \wrt a preorder $ \RelT \subseteq \procT^2 $ that is a bisimulation iff $ \exists \IRel \logdot \left( \forall S \logdot \left( S, \Encoding{S} \right) \in \IRel \right) \wedge \RelT = \Red{\IRel}{\procT} $\linebreak $ \wedge \left( \forall S, T \logdot \left( S, T
        \right) \in \IRel \rightarrow \left( \Encoding{S}, T \right) \in \RelT \right) \wedge \IRel $ is a preorder and a bisimulation.
\end{lemma}

\noindent
We obtain the same result if we replace operational correspondence by strong operational correspondence and bisimulation by strong bisimulation.

Weak and strong bisimilarity are often considered as \textit{the} standard reference relations for calculi like the $ \pi $-calculus. Thus the above result imposes an important property for the comparison of languages. If bisimilarity is the standard reference relation, \ie if we usually do not record differences between terms that cannot be observed by bisimilarity, then an encoding that ensures that source terms and their translations are bisimilar strongly validates the claim that the target language is at least as expressive as the source language. Nonetheless, comparisons of different languages are very often considered only modulo weak operational correspondence and not operational correspondence. As discussed in \cite{parrowCoupled92,cspToCcs15}, relating source terms and their literal translations by a bisimulation does not allow for intermediate states, \ie states that occur in simulations of source term steps and thus intuitively are in between two source term translations but are not related to source terms themselves. Intermediate states result from partial commitments. If a source term can evolve to one of three different derivatives, operational correspondence (in all variants) ensures that the translation has the same possible evolutions. But operational correspondence requires that the decision on which of the three possibilities is chosen is done in a single step. Weak operational correspondence allows for partial commitments, where a first step may rule out one possibility but not decide on one of the remaining two.
Thus weak operational correspondence is much more flexible and allows to encode source term concepts that have no direct counterpart in the target.

Obtaining a result similar to Lemma~\ref{lem:operationalCorrespondence} for weak operational correspondence is not that easy. Again this criterion is linked to a simulation condition on relations between source and target terms up to $ \RelT $, but weak operational correspondence does not directly map to a well-known kind of simulation relation. It is linked to a simulation relation that is in between coupled similarity and bisimilarity.
We call it correspondence similarity.

\begin{definition}[Correspondence Simulation]
	A relation $ \Rel $ is a \emph{(weak reduction) correspondence simulation} if for each $ (P, Q) \in \Rel $:
	\begin{compactitem}
		\item $ P \steps P' $ implies $ \exists Q' \logdot Q \steps Q' \wedge (P', Q') \in \Rel $
		\item $ Q \steps Q' $ implies $ \exists P'', Q'' \logdot P \steps P'' \wedge Q' \steps Q'' \wedge (P'', Q'') \in \Rel $
	\end{compactitem}
	Two terms are \emph{correspondence similar} if a correspondence simulation relates them.
\end{definition}

\noindent
Just as coupled similarity, correspondence similarity allows for intermediate states that result from partial commitments,
but in contrast to coupled similarity these intermediate states are not necessarily covered in the relation.
Correspondence similarity is obviously strictly weaker than
bisimilarity, but it implies coupled similarity.

\begin{lemma}\label{lem:correspondance-coupled}
	For each correspondence simulation $ \Rel $ there exists a coupled simulation $ \Rel' $ such that $ \forall \left( P, Q \right) \in \Rel \logdot \left( P, Q \right), \left( Q, P \right) \in \Rel' $.
\end{lemma}

Correspondence simulation is linked to weak operational correspondence in the same way as bisimilarity is linked to operational correspondence.

\begin{lemma}[Weak Operational Correspondence]
	$ \encoding : \procS \to \procT $ is weakly operat.\ corresp.\ \wrt a preorder $ \RelT \subseteq \procT^2 $ that is a correspondence simulation iff $ \exists \IRel \logdot \left( \forall S \logdot \left( S, \Encoding{S} \right) \in \IRel \right) \wedge \RelT = \Red{\IRel}{\procT} $\\ $ \wedge \; \left( \forall S, T \logdot \left( S, T \right) \in \IRel \rightarrow \left( \Encoding{S}, T \right) \in \RelT \right) \wedge \IRel $ is a preorder and a correspondence simulation.
	\label{lem:WOC}
\end{lemma}

\noindent
Accordingly, weak operational correspondence ensures that source terms and their literal translations are correspondence similar and thus coupled similar.

Correspondence similarity and coupled similarity are weaker than bisimilarity. Nevertheless, proving
that a relation is a correspondence simulation and, even more, showing that a particular pair of
terms is contained in a correspondence simulation, can be more difficult than it is in the case of
bisimulation. Fortunately, encodings that satisfy only weak operational correspondence---and introduce
partial commitments---often do so \wrt a variant of bisimilarity. As example consider the
de-centralised encoding of \cite{cspToCcs15}. It translates from CSP into asynchronous CCS with name
passing and matching. \cite{cspToCcs15} proves that this encoding is operational corresponding \wrt
a target term preorder $ \RelT $ that is a weak reduction bisimulation. Thus, by
Lemma~\ref{lem:WOC}, the encoding ensures that source terms and their literal translations are
correspondence similar, and thus coupled similar.

\section{Combining Encodability Criteria}
\label{sec:combining}

As done in \cite{Gorla10a}, often several different criteria are combined to ensure the quality
of an encoding. Of course we have to ensure that the criteria we want to combine do not contradict
each other and thus trivially rule out any kind of encoding. Moreover, the combination of criteria
might lead to unexpected side effects, such that their combined effect on the quality of encodings is
no longer obvious or clear. One major motivation of our desire to analyse encodability criteria is
to be able to formally compare them and analyse side effects that result from their combinations.

In the previous section we derive iff-results linking a single criterion with the existence
of a relation between source and target terms satisfying specific conditions. Of course we can trivially
combine two such results by considering two different source-target relations on the right-hand side. But
this way side effects that results from the combination of the criteria remain hidden. Instead we
want to combine the criteria into conditions of a single source-target relation. Therefore we need
to find a witness, \ie a relation that satisfies the conditions of both relations.

\subsection{Divergence Reflection and Success Sensitiveness}
\label{sec:divergenceAndSuccess}

The combinations of criteria defined on the pairs of $ \IRel $---such as
the preservation, reflection, or respect of some predicate---are easy to analyse. Obviously an
encoding reflects divergence and respects success iff there exists a relation $ \IRel $ that relates
at least each source term to its literal translation and both reflects divergence and respects
success.

\begin{lemma}
	\label{lem:DivReflSuccResp}
	Assume $ \langS $ and $ \langT $ each define a predicate $ \hasSuccess: \proc \to \bool $.
	$ \encoding : \procS \to \procT $ reflects divergence and respects success iff $ \exists \IRel\logdot \left( \forall S\logdot \left( S, \Encoding{S} \right) \in \IRel \right) \wedge \IRel $ reflects divergence and respects success.
\end{lemma}

\noindent
The $\Leftarrow$-direction of Lemma~\ref{lem:DivReflSuccResp} is an immediate corollary of Lemmas~\ref{lem:divergence} and~\ref{lem:success}.
For the other direction we obtain from these lemmata two relations that satisfy the condition $ \cond \deff \forall S\logdot \left( S, \Encoding{S} \right) \in \IRel $ and of which one
reflects divergence and the other respects success. We have to combine these two relations into a
single relation that satisfies all three conditions. If the latter two conditions are defined on the
pairs of the respective relations, this is always possible. The reason is that the condition $ \cond
$ ensures that we can use $ \Set{ \left( S, \Encoding{S} \right) \mid S \in \procS } $ as a witness
for both relations and thus as a witness for their combined effect. More precisely, if there are two
relations that both satisfy $ \cond $ and each satisfies a predicate about the pairs of the
respective relation, then there exists a single relation, namely $ \Set{ \left( S, \Encoding{S} \right) \mid
  S \in \procS } $, that satisfies all three conditions.

\begin{lemma}
	Let $ \Rel_1, \Rel_2 \subseteq \left( \procS \uplus \procT \right)^2 $ and the predicates $ \predA, \predB $ be such that $ \forall i \in \Set{ 1, 2 }\logdot \forall S\logdot \left( S, \Encoding{S} \right) \in \Rel_i $ and $ \forall i \in \Set{ 1, 2 }\logdot \forall \left( P, Q \right) \in \Rel_i\logdot \PredI{i}{\left( P, Q \right)} $.
	Then there exists a relation $ \IRel \subseteq \left( \procS \uplus \procT \right)^2 $ such that $ \forall S\logdot \left( S, \Encoding{S} \right) \in \IRel $ and $ \forall i \in \Set{ 1, 2 }\logdot \forall \left( P, Q \right) \in \IRel\logdot \PredI{i}{\left( P, Q \right)} $.
\end{lemma}

\subsection{Adding Operational Correspondence}
\label{sec:addingOperationalCorrespondence}

Gorla \cite{Gorla10a} combines five criteria to define `good' encodings. Three of these---the `semantical' ones---we
considered in Section~\ref{analysing}:
 weak operational correspondence (called `operational correspondence'
in \cite{Gorla10a}) \wrt a relation $ \RelT $, success sensitiveness, and divergence reflection.
Gorla assumes that, `for the sake of coherence' as he claims, the relation $ \RelT $
never relates two process $ T_P $ and $ T_Q $ such that $ T_P \reachSuccess $ and $ T_Q\not\,\reachSuccess $,
\ie $ \RelT $ has to respect (reachability of) success. This allows us to find a witness relation to combine the
effect of weak operational correspondence and success sensitiveness.
Our iff-result for weak operational correspondence requires that this relation is a preorder, has to relate source terms with their literal
translations, and satisfies $ \RelT = \Red{\IRel}{\procT} $. Because of that, a minimal witness is $
\trans{\refl{\Set{\left( S, \Encoding{S} \right) \mid S \in \procS} \cup \RelT}} $. This
witness also satisfies $\forall S, T \logdot \left( S, T \right) \in \IRel \rightarrow \left( \Encoding{S}, T \right) \in \RelT $. Without the
condition that $ \RelT $ respects success---or another suitable assumption---we cannot ensure that $
\trans{\refl{\Set{\left( S, \Encoding{S} \right) \mid S \in \procS} \cup \RelT}} $ respects success
and thus we find no witness for the combination of the respective conditions.

\begin{lemma}\label{lem:oper. corr. and success}
	Assume $ \langS, \langT $ each define a predicate $ \hasSuccess: \proc \to \bool $.
	An encoding $ \encoding : \procS \to \procT $
        is success sensitive and weakly operational corresponding
        \wrt a preorder $ \RelT \subseteq \procT^2 $ that is a success respecting
        correspondence simulation iff $ \exists \IRel \logdot \left( \forall S \logdot \left( S,
        \Encoding{S} \right) \in \IRel \right) \wedge \RelT = \Red{\IRel}{\procT} \wedge  \; \IRel $
        respects success $\mbox{}\wedge \left( \forall S, T \logdot \left( S, T \right) \in \IRel
        \rightarrow \left( \Encoding{S}, T \right) \in \RelT \right) $ $ \wedge \; \IRel $
        is a preorder and a correspondence simulation.
\end{lemma}

\noindent
We obtain a similar result if we replace weak operational correspondence and correspondence
simulation by operational correspondence and bisimulation. Similarly, we can replace weak operational
correspondence, correspondence simulation, and the predicate $ \reachSuccess $ in the definition of
success sensitiveness by strong operational correspondence, strong bisimulation, and $ \hasSuccess $.
Also, in all variants of the above result we can add on the left-hand side that the encoding as well as
$ \RelT $ (weakly) respect barbs iff we add on the right-hand side that
$ \IRel $ (weakly) respects barbs.

As an example we can consider---once more---the de-centralised encoding from CSP into a variant of CCS of \cite{cspToCcs15}. Additionally to operational correspondence \wrt a preorder $ \RelT $ that is a bisimulation, \cite{cspToCcs15} proves that this encoding satisfies success sensitiveness, divergence reflection, barb sensitiveness---\wrt standard CSP barbs in the source and a notion of translated barbs in the target---and preservation of distributability (a criterion defined in \cite{petersNestmannGoltz13}). $ \RelT $ respects success and weakly respects barbs (but does not reflect divergence). Thus the encoding ensures that source terms and their literal translations are correspondence similar and thus coupled similar \wrt a relation that respects success and weakly respects barbs.

Success sensitiveness significantly strengthens the requirements on a simulation relation like correspondence simulation or bisimulation. Thus the combined effect---a success respecting correspondence simulation---is stronger than the effects of both criteria considered in isolation---a correspondence simulation and a success respecting relation.
Accordingly, the framework of Gorla in \cite{Gorla10a} ensures that (among other conditions) source terms and their literal translations are correspondence similar \wrt a success respecting relation and thus---to refer to a more established simulation relation---are coupled similar \wrt a success respecting relation.

In \cite{Gorla10a} there is no such condition that links $ \RelT $ and divergence
reflection. Requiring that $ \RelT $ reflects divergence would \eg exclude weak bisimulation. Since
this relation is often referred to as \textit{the} standard relation for calculi as the $ \pi
$-calculus, excluding it would be too strict a requirement. As a consequence, the criteria in
\cite{Gorla10a} do not allow to combine the effects of weak operational correspondence and
divergence reflection into a single relation, as done for success sensitiveness.
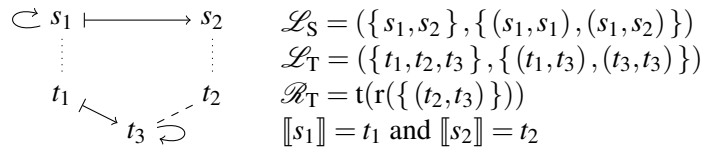
\begin{figure}[t]
	\centering
	\begin{tabular}{cc}
		\begin{tikzpicture}[]
			\node (s1) at (0, 1.5)	{$ s_1 $};
			\node (s2) at (2, 1.5)	{$ s_2 $};
			\node (t1) at (0, 0.5)	{$ t_1 $};
			\node (t2) at (2, 0.5)	{$ t_2 $};
			\node (t3) at (1, 0)		{$ t_3 $};
			
			\path[|->] (s1) edge (s2);
			\path[|->] (s1) [loop left] edge ();
			\path[|->] (t1) edge (t3);
			\path[|->] (t3) [loop right] edge ();
			
			\path[dotted] (s1) edge (t1);
			\path[dotted] (s2) edge (t2);
			
			\path[dashed] (t2) edge (t3);
		\end{tikzpicture}
		&
		\begin{tabular}[b]{l}
			$ \langS = \left( \Set{ s_1, s_2 }, \Set{ \left( s_1, s_1 \right), \left( s_1, s_2 \right) } \right) $\\
			$ \langT = \left( \Set{ t_1, t_2, t_3 }, \Set{ \left( t_1, t_3 \right), \left( t_3, t_3 \right) } \right) $\\
			$ \RelT = \trans{\refl{\Set{ \left( t_2, t_3 \right) }}} $\\
			$ \Encoding{s_1} = t_1 $ and $ \Encoding{s_2} = t_2 $ \vspace{0.1em}
		\end{tabular}
	\end{tabular}
	\caption{Encoding satisfying operational correspondence and divergence reflection}
	\label{fig:counterexOCDR}
\end{figure}
Consider the following counterexample, visualised in Figure~\ref{fig:counterexOCDR}. Obviously the
encoding---indicated by the dotted line---satisfies operational correspondence \wrt $ \RelT $---indicated by the dashed line---and reflects divergence. But to relate
$ s_1 $ and its literal translation $ \Encoding{s_1} = t_1 $ by a correspondence simulation, we have
to simulate the step $ s_1 \stepS s_2 $. Therefore we need either the pair $ \left( s_2, t_3 \right)
$---which can be obtained by including $ \RelT $ in $ \IRel $---or the pair $ \left( s_2, t_1
\right) $, but in either case the respective source-target relation does not reflect divergence.
Thus in general an encoding that satisfies the criteria of \cite{Gorla10a} induces a source-target relation that is a correspondence simulation that only partially reflects divergence.

Of course particular encodings might satisfy stronger requirements than enforced by the minimal setting in \cite{Gorla10a}. If the encoding is operational corresponding \wrt a relation that reflects divergence, we can combine the effects of these two criteria in one relation. Accordingly, if an encoding reflects divergence, respects success, and satisfies operational correspondence \wrt a preorder that is a success respecting and divergence reflecting bisimulation, we can combine the conditions of all three relations
as in the following lemma.

\begin{lemma}
	Assume $ \langS $ and $ \langT $ each define a predicate $ \hasSuccess: \proc \to \bool $.
	An encoding $ \encoding : \procS \to \procT $ reflects divergence, respects success, and is operational
        corresponding \wrt a preorder $ \RelT \subseteq \procT^2 $ that is a success respecting and divergence
        reflecting bisimulation iff $ \exists \IRel \logdot \left( \forall S \logdot \left(
        S, \Encoding{S} \right) \in \IRel \right) \wedge \RelT = \Red{\IRel}{\procT} \wedge \left(
        \forall S, T \logdot \left( S, T \right) \in \IRel \rightarrow \left( \Encoding{S}, T
        \right) \in \RelT \right) $ $ \wedge \; \IRel $ reflects divergence, respects success, and
        is a preorder and a bisimulation.
\end{lemma}

\noindent
Again we obtain similar results for weak operational correspondence and correspondence simulation as well as for strong operational correspondence, strong bisimulation, and $ \hasSuccess $.

\subsection{Full Abstraction and Operational Correspondence}
\label{sec:fullAbstractionAndOperationalCorrespondence}

Before the framework in \cite{Gorla10a} was proposed, often a combination of full abstraction and
operational correspondence was used. For simplicity we switch to source-target relations that relate
source terms and their literal translations in both directions and assume that $ \RelS $ and $ \RelT
$ are equivalences in the following.
Then a witness for the effect of operational correspondence is $ \trans{\refl{\Set{\left( S,
      \Encoding{S} \right), \left( \Encoding{S}, S \right) \mid S \in \procS} \cup \RelT}} $. Since
this relation is transitive, it indeed suffices as witness to combine the effects of full
abstraction and operational correspondence. The only obstacle left is that, to cover the effect of
full abstraction, the source-target relation should also include $ \RelS $. Fortunately we do not have to include $ \RelS $ by construction, because its inclusion is ensured by full abstraction and the inclusion of $ \RelT $.
For every encoding $ \encoding $ that is fully abstract \wrt $ \RelS $ and $ \RelT $ and for all transitive relations $ \IRel $ that relate at least all source terms to their literal translations in both directions, $ \IRel $ contains $ \RelS $ iff the restriction of $ \IRel $ to encoded source terms contains the restriction of $ \RelT $ to encoded sources.

\begin{lemma}
	Let $ \encoding : \procS \to \procT $ be an encoding that is fully abstract \wrt $ \RelS \subseteq \procS^2 $ and $ \RelT \subseteq \procT^2 $ and let $ \IRel \subseteq \left( \procS \uplus \procT \right)^2 $ be transitive such that $ \forall S \logdot \left( S, \Encoding{S} \right), \left( \Encoding{S}, S \right) \in \IRel $. Then $ \RelS = \Red{\IRel}{\procS} $ iff $ \forall S_1, S_2\logdot \left( \Encoding{S_1}, \Encoding{S_2} \right) \in \RelT \leftrightarrow \left( \Encoding{S_1}, \Encoding{S_2} \right) \in \IRel $.
\end{lemma}

Because of that an encoding is fully abstract \wrt $ \RelS $ and $ \RelT $ and operational corresponding \wrt a bisimulation $ \RelT $ iff there exists a transitive bisimulation that relates source terms and their literal translations in both directions and contains $ \RelS $ and $ \RelT $.

\begin{lemma}
	Let $ \RelS \subseteq \procS^2 $ and $ \RelT \subseteq \procT^2 $ be equivalences. An encoding $ \encoding : \procS \to \procT $ is fully abstract \wrt $ \RelS $ and $ \RelT $ and operational corresponding \wrt $ \RelT $ and $ \RelT $ is a bisimulation iff $ \exists \IRel \logdot $  \linebreak $ \left( \forall S \logdot \left( S, \Encoding{S} \right), \left( \Encoding{S}, S \right) \in \IRel \right) \wedge \RelS = \Red{\IRel}{\procS} \wedge \RelT = \Red{\IRel}{\procT} \wedge \IRel $ is a transitive bisimulation.
\end{lemma}

So what do we gain by combining the two criteria, that we do not obtain from each of them in isolation? In comparison to our iff-result for operational correspondence we add only the condition that $ \RelS = \Red{\IRel}{\procS} $. As a consequence $ \RelS $ has to be a bisimulation.

Full abstraction ensures that $ \RelS $ and $ \RelT $ have the same basic properties. For
example, if we either consider surjective encodings ($ \forall T\logdot \exists S\logdot T = \Encoding{S} $) or restrict $ \RelT $ to encoded source terms ($ \{ \left( T_1, T_2 \right) \mid \exists S_1, S_2\logdot T_1 = \Encoding{S_1} \wedge T_2 = \Encoding{S_2} \wedge \left( T_1, T_2 \right) \in \RelT \} $), then $ \RelS $ is reflexive iff $ \RelT $ is reflexive,
and similarly for symmetry and transitivity.
But properties such as being a bisimulation are not respected by full abstraction on its own.
\begin{figure}[t]
	\centering
	\begin{tabular}{cc}
		\begin{tikzpicture}[]
			\node (s1) at (0, 1)	{$ s_1 $};
			\node (s2) at (2, 1)	{$ s_2 $};
			\node (s3) at (4, 1)	{$ s_3 $};
			\node (t1) at (0, 0)	{$ t_1 $};
			\node (t2) at (2, 0)	{$ t_2 $};
			\node (t3) at (4, 0)	{$ t_3 $};
			
			\path[|->] (t2) edge (t3);
			
			\path[dotted] (s1) edge (t1);
			\path[dotted] (s2) edge (t2);
			\path[dotted] (s3) edge (t3);
			
			\path[dashed] (s1) [bend left] edge (s2);
			\path[dashed] (t1) [bend right] edge (t2);
		\end{tikzpicture}
		&
		\begin{tabular}[b]{l}
			$ \langS = \left( \Set{ s_1, s_2 , s_3 }, \emptyset \right) $\\
			$ \langT = \left( \Set{ t_1, t_2 , t_3 }, \Set{ \left( t_2, t_3 \right) } \right) $\\
			$ \RelS = \trans{\sym{\refl{\Set{ \left( s_1, s_2 \right) }}}} $ and $ \RelT = \trans{\sym{\refl{\Set{ \left( t_1, t_2 \right) }}}} $\\
			$ \Encoding{s_1} = t_1 $, $ \Encoding{s_2} = t_2 $, and $ \Encoding{s_3} = t_3 $ \vspace{0.1em}
		\end{tabular}
	\end{tabular}
	\caption{Encoding satisfying full abstraction but not operational correspondence}
	\label{fig:counterexFAOC}
\end{figure}
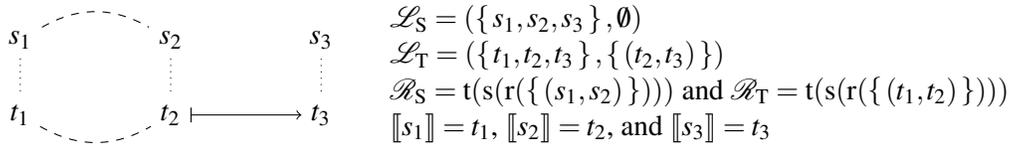
As counterexample consider the fully abstract but not operational corresponding encoding in Figure~\ref{fig:counterexFAOC}. Here $ \RelS $ is a bisimulation but $ \RelT $ is not. By removing the arrow $ t_2 \stepT t_3 $ from the target and adding it to the source $ s_2 \stepS s_3 $, the encoding remains fully abstract and $ \RelT $ becomes a bisimulation but $ \RelS $ loses this property.
Operational correspondence does not refer to a source relation $ \RelS $ and thus does not enforce any properties on this relation.
But combining full abstraction with operational correspondence \wrt a bisimulation $ \RelT $ enforces $ \RelS $ to be a bisimulation.

\begin{lemma}
	Let $ \RelS \subseteq \procS^2 $ and $ \RelT \subseteq \procT^2 $ be equivalences. If an encoding $ \encoding : \procS \to \procT $ is fully abstract \wrt $ \RelS $ and $ \RelT $ and operational corresponding \wrt $ \RelT $ and $ \RelT $ is a bisimulation then $ \RelS $ is a bisimulation.
\end{lemma}

To conclude that $ \RelS $ is a bisimulation iff $ \RelT $ is a bisimulation, we have to get rid of pairs in $ \RelT $ that do not result from pairs of encoded source terms and their derivatives, because operational correspondence provides no information about such pairs. The simplest way to do so, is to assume a surjective encoding.

\begin{lemma}
	Let $ \RelS \subseteq \procS^2 $ and $ \RelT \subseteq \procT^2 $ be equivalences. If an encoding $ \encoding : \procS \to \procT $ is surjective ($ \forall T\logdot \exists S\logdot T = \Encoding{S} $), fully abstract \wrt $ \RelS $ and $ \RelT $, and operational corresponding \wrt $ \RelT $ then:
	\begin{center}
		$ \RelS $ is a bisimulation iff $ \RelT $ is a bisimulation
	\end{center}
\end{lemma}

\section{Conclusions}

Within this paper we provide a number of results about different encodability criteria. In particular:
\begin{compactitem}
	\item We analyse divergence reflection, barb sensitiveness, success sensitiveness, full abstraction, and operational correspondence as well as several combinations of these criteria.
	\item We prove that different variants of operational correspondence correlate with different kinds of simulation relations from coupled similarity to strong bisimilarity.
	\item We define a new kind of simulation relation---correspondence similarity---that completely covers the effect of weak operational correspondence as proposed in \cite{Gorla10a}.
	\item We relate the combination of success sensitiveness and operational correspondence \wrt a bisimulation with the existence of a success respecting bisimulation between source terms and their literal translations.
	\item We show that for surjective encodings the combination of full abstraction \wrt $ \RelS $ and $ \RelT $ and operational correspondence \wrt $ \RelT $ implies that $ \RelS $ is a bisimulation iff $ \RelT $ is a bisimulation.
\end{compactitem}

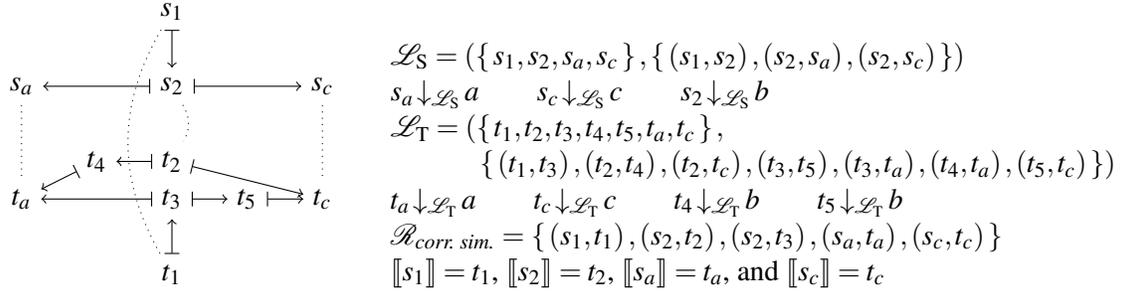
\begin{figure}[t]
	\centering
	\begin{tabular}{cc}
		\begin{tikzpicture}[]
			\node (s1) at (2, 3.5)	{$ s_1 $};
			\node (s2) at (2, 2.5)	{$ s_2 $};
			\node (sa) at (0, 2.5)	{$ s_a $};
			\node (sc) at (4, 2.5)	{$ s_c $};
			\node (t1) at (2, 0)		{$ t_1 $};
			\node (t2) at (2, 1.5)	{$ t_2 $};
			\node (t3) at (2, 1)		{$ t_3 $};
			\node (t4) at (1, 1.5)	{$ t_4 $};
			\node (t5) at (3, 1)		{$ t_5 $};
			\node (ta) at (0, 1)		{$ t_a $};
			\node (tc) at (4, 1)		{$ t_c $};
			
			\path[|->] (s1) edge (s2);
			\path[|->] (s2) edge (sa);
			\path[|->] (s2) edge (sc);
			\path[|->] (t1) edge (t3);
			\path[|->] (t2) edge (t4);
			\path[|->] (t2) edge (tc);
			\path[|->] (t3) edge (t5);
			\path[|->] (t3) edge (ta);
			\path[|->] (t4) edge (ta);
			\path[|->] (t5) edge (tc);
			
			\path[dotted] (s1) [bend right]	edge (t1);
			\path[dotted] (s2) [bend left]	edge (t2);
			\path[dotted] (sa)				edge (ta);
			\path[dotted] (sc)				edge (tc);
		\end{tikzpicture}
		&
		\begin{tabular}[b]{l}
			$ \langS = \left( \Set{ s_1, s_2, s_a, s_c }, \Set{ \left( s_1, s_2 \right), \left( s_2, s_a \right), \left( s_2, s_c \right) } \right) $\\
                        $ \HasBarbL{\langS}{s_a}{a} \qquad \HasBarbL{\langS}{s_c}{c} \qquad \HasBarbL{\langS}{s_2}{b} $ \\
			$ \langT = ( \Set{ t_1, t_2, t_3, t_4, t_5, t_a, t_c }, $\\
			$ \phantom{\langT = (} \Set{ \left( t_1, t_3 \right), \left( t_2, t_4 \right), \left( t_2, t_c \right), \left( t_3, t_5 \right), \left( t_3, t_a \right), \left( t_4, t_a \right), \left( t_5, t_c \right) } ) $\\
                        $ \HasBarbL{\langT}{t_a}{a} \qquad \HasBarbL{\langT}{t_c}{c} \qquad \HasBarbL{\langT}{t_4}{b} \qquad \HasBarbL{\langT}{t_5}{b} $ \\ 
			$ \mathcal{R}_{\textit{corr.\ sim.}} = \Set{ \left( s_1, t_1 \right), \left( s_2, t_2 \right), \left( s_2, t_3 \right), \left( s_a, t_a \right), \left( s_c, t_c \right) } $\\
			$ \Encoding{s_1} = t_1 $, $ \Encoding{s_2} = t_2 $, $ \Encoding{s_a} = t_a $, and $ \Encoding{s_c} = t_c $ \vspace{0.1em}
		\end{tabular}
	\end{tabular}
	\caption{Encoding not satisfying weak operational correspondence \wrt to a correspondence
          simulation that weakly respects barbs, even though $\mathcal{R}_{\textit{corr.\ sim.}}$ is a
          weakly barb respecting correspondence simulation relating each source term with its encoding}
	\label{fig:counterexMusing}
\end{figure}

In \cite{vG12} a quality criterion for encodings was proposed that requires the translation $\Encoding{S}$ of a source term $S$ to be related to $S$ according to a behavioural equivalence or preorder defined on a domain of interpretation (such as labelled transition systems or reduction-based transition systems with barbs) that applies to both languages. This behavioural relation has to be chosen with care and should be meaningful for the application at hand.
Possible choices include strong and weak barbed bisimilarity, barbed weak coupled simulation equivalence, or (in between) our new correspondence preorder. Iff-results---as the results above---relate these
instances of the criterion discussed in \cite{vG12} with other encodability criteria.
In particular, by the results of Section~\ref{sec:analysing}, if an encoding satisfies
the criterion of \cite{vG12} \wrt (weak) bisimilarity, then it also satisfies operational 
correspondence \wrt (weak) bisimilarity.\footnote{And by the results of Section 4 the bisimulation
  may be required to (weakly) respect barbs at both sides of the implication.}
Moreover, if an encoding satisfies the criterion of \cite{vG12} \wrt correspondence similarity, then
it also satisfies weak operational correspondence \wrt coupled similarity.\footnote{We
 may not conclude that it also satisfies weak operational correspondence \wrt correspondence
 similarity, at least not when also weakly respecting barbs. A counterexample can be found in
 Figure~\ref{fig:counterexMusing}.  There we have a weakly barb respecting correspondence simulation
 relating $s_2$ with $t_2$ and $s_2$ with $t_3$, but, due to the asymmetric nature of correspondence
 simulations, there is no weakly barb respecting correspondence simulation relating $t_2$ and $t_3$.\footnotemark{}
 By Lemma~\ref{lem:correspondance-coupled} the terms $t_2$ and $t_3$ are weakly barb respecting
 coupled similar, however.}
 \pagebreak[3]
 \footnotetext{This example does not contradict the weakly barbed variant of
 Lemma~\ref{lem:oper. corr. and  success}, for its right-hand side does not hold. Namely, the condition
 $\forall S \logdot \left( S, \Encoding{S} \right) \in \IRel$ forces $(s_1,t_1) \in \IRel$, and
 thus, since $\IRel$ is a correspondence simulation, also $(s_2,t) \in \IRel$ for some $t\in \{t_1,t_3,t_5,t_a,t_c\}$.
 As $s_2$ weakly respects barbs $a$ and $c$, so must $t$, yielding $t\in \{t_1,t_3\}$. The requirement
 $\left( S, T \right) \in \IRel \rightarrow \left( \Encoding{S}, T \right) \in \RelT$
 yields $(t_2,t) \in \IRel$, but there exists no correspondence simulation containing this pair.}
 
The above results may leave the impression that we try to replace common encodability criteria by conditions on relations between source terms and their translations. That is not the case. But we provide alternative ways to prove different criteria.
An example of how the above results can be used to reason about the quality of an encoding are the
two encodings of \cite{cspToCcs15}. That paper analyses ways to encode the CSP synchronisation
mechanism following an approach of \cite{parrowCoupled92}. The latter shows that a central encoding of a similar synchronisation mechanism ensures that source terms and their translations behave bisimilar, whereas a decentralised encoding only ensures coupled similarity.
Proving coupled similarity can be more difficult than proving bisimilarity. Here our results allow to decrease the proof burden. With Lemma~\ref{lem:WOC} we can conclude from weak operational correspondence \wrt a bisimulation that source terms and their literal translations are coupled similar, without having to deal with coupled similarity directly. This way we have to deal with the difficult partial commitments, which are introduced by the de-central implementation, only in operational correspondence and not when relating target terms.

In retrospective, mapping encodability criteria on requirements of a relation between source and target terms seems quite natural. Indeed the main challenge of the above presented iff-results was not in proving them but in finding the exact matches between variants of the considered criteria and requirements on the relation. As a consequence we had to define a new kind of simulation relation to capture the version of operational correspondence used in \cite{Gorla10a}.

We do not claim that it is always simple to obtain iff-results as presented in this paper or that we provide a strategy to obtain  such results. Instead we claim that proving such results formally captures the effect of a criterion on the quality of an encoding function and thus\!\!
\begin{inparaenum}[(1)]
	\item helps us to understand a criterion,
	\item allows to identify unexpectedly strict or weak criteria
	\item allows to compare (sets of) criteria, and
	\item allows to analyse the side effects that result from the combination of criteria.
\end{inparaenum}
Analysing criteria this way is not necessarily straightforward.
To illustrate this, consider the requirement on the preservation of the (degree of) distribution of a process (preservation of distributability). In the context of asynchronous distributed systems this requirement is very important.

Several attempts to capture it were proposed in the literature.\pagebreak[3]
At least for the $ \pi $-calculus, the most prominent candidate is the homomorphic translation of the parallel operator ($ \Encoding{P \mid Q} = \Encoding{P} \mid \Encoding{Q}) $ as used in \cite{Palamidessi03}.
\cite{petersNestmannGoltz13} shows that this criterion is too strict by providing an encoding that
preserves distributability but does not translate the parallel operator homomorphically. Instead
this encoding is compositional. Compositionality requires that the translation of an operator is the
same for all occurrences of that operator in a term, \ie it can be captured by a
context. Compositionality is significantly weaker than the homomorphic translation of the parallel
operator but also forbids the introduction of global coordinators.\linebreak
However, to ensure the preservation of distributability, this criterion is too weak.
\cite{petersNestmannGoltz13} claims to provide a suitable criterion for the degree of distributability, but without a formal way to reason about the effect of encodability criteria, there is no way to formally prove such a claim. Thus \cite{petersNestmannGoltz13} can only provide arguments and illustrations. The inability to formally prove it was one of the original motivations for the present work.
Unfortunately, analysing the three criteria compositionality, the homomorphic translation of the parallel operator, and the preservation of distributability is not an easy task.

Compositionality obviously implies some kind of congruence property on encoded source term contexts, but it is not obvious how to turn this observation into an iff-result. To map the homomorphic translation of the parallel operator on conditions on a relation between source and target terms, is even more difficult. This criterion clearly implies some strong properties on such a relation, but it is not clear which condition implies the homomorphic translation of the parallel operator.
Because of that, we cannot completely capture the effect of this criterion on the quality of an
encoding. This explains why this criterion was originally accepted as a criterion for the
preservation of distributability. It is not easy to capture the cases for which it is too strict.
The criterion for the preservation of distributability proposed in \cite{petersNestmannGoltz13} can
intuitively be understood as a concurrency respecting variant of operational correspondence. It not
only requires that the source term behaviour is preserved and reflected, but also that the
simulations of independent steps are independent. Thus analysing this criterion appears to require
some kind of simulation relations that not only consider interleaving semantics. We leave the
analysis of these criteria to further research.

All claims in this paper have been proved using the interactive theorem prover Isabelle/HOL.
For this purpose, a rich theory of encodability criteria was implemented. Since we do not force any assumptions on process calculi except that they consist of a set of processes, \ie a type $ \proc $, and a reduction relation, \ie a relation of type $ \proc^2 = \proc \times \proc $, this theory can be used to formally reason in Isabelle about encodings for all kinds of source and target languages.
A number of well-known process calculi including the $ \pi $-calculus can for instance be represented in the Psi-calculi framework \cite{psiFramework09}. Thus there are Isabelle implementations of well-known process calculi that can directly be combined with our Isabelle implementation to formally reason about encodings between such calculi.

\providecommand{\thisvolume}[2][]{this volume of EPTCS}
\def\opa{}
\bibliography{encodings,qualityCriteria}

\begin{thebibliography}{10}
\providecommand{\bibitemdeclare}[2]{}
\providecommand{\surnamestart}{}
\providecommand{\surnameend}{}
\providecommand{\urlprefix}{Available at }
\providecommand{\url}[1]{\texttt{#1}}
\providecommand{\href}[2]{\texttt{#2}}
\providecommand{\urlalt}[2]{\href{#1}{#2}}
\providecommand{\doi}[1]{doi:\urlalt{http://dx.doi.org/#1}{#1}}
\providecommand{\bibinfo}[2]{#2}

\bibitemdeclare{inproceedings}{psiFramework09}
\bibitem{psiFramework09}
\bibinfo{author}{J.~\surnamestart Bengtson\surnameend},
  \bibinfo{author}{M.~\surnamestart Johansson\surnameend},
  \bibinfo{author}{J.~\surnamestart Parrow\surnameend} \&
  \bibinfo{author}{B.~\surnamestart Victor\surnameend} (\bibinfo{year}{2009}):
  \emph{\bibinfo{title}{{Psi-calculi: Mobile Processes, Nominal Data, and
  Logic}}}.
\newblock In: {\sl \bibinfo{booktitle}{Proceedings of LICS}},
  \bibinfo{organization}{IEEE}, pp. \bibinfo{pages}{39--48},
  \doi{10.1109/LICS.2009.20}.

\bibitemdeclare{article}{BusiGZ09}
\bibitem{BusiGZ09}
\bibinfo{author}{N.~\surnamestart Busi\surnameend},
  \bibinfo{author}{M.~\surnamestart Gabbrielli\surnameend} \&
  \bibinfo{author}{G.~\surnamestart Zavattaro\surnameend}
  (\bibinfo{year}{2009}): \emph{\bibinfo{title}{On the expressive power of
  recursion, replication and iteration in process calculi}}.
\newblock {\sl \bibinfo{journal}{Mathematical Structures in Computer Science}}
  \bibinfo{volume}{19}(\bibinfo{number}{6}), pp. \bibinfo{pages}{1191--1222},
  \doi{10.1017/S096012950999017X}.

\bibitemdeclare{article}{CCAV08}
\bibitem{CCAV08}
\bibinfo{author}{D.~\surnamestart Cacciagrano\surnameend},
  \bibinfo{author}{F.~\surnamestart Corradini\surnameend},
  \bibinfo{author}{J.~\surnamestart Aranda\surnameend} \& \bibinfo{author}{F.D.
  \surnamestart Valencia\surnameend} (\bibinfo{year}{2008}):
  \emph{\bibinfo{title}{{Linearity, Persistence and Testing Semantics in the
  Asynchronous Pi-Calculus}}}.
\newblock {\sl \bibinfo{journal}{Electronic Notes in Theoretical Computer
  Science}} \bibinfo{volume}{194}(\bibinfo{number}{2}), pp.
  \bibinfo{pages}{59--84}, \doi{10.1016/j.entcs.2007.11.006}.

\bibitemdeclare{article}{cardelliGordon00}
\bibitem{cardelliGordon00}
\bibinfo{author}{L.~\surnamestart Cardelli\surnameend} \&
  \bibinfo{author}{A.~D. \surnamestart Gordon\surnameend}
  (\bibinfo{year}{2000}): \emph{\bibinfo{title}{{Mobile ambients}}}.
\newblock {\sl \bibinfo{journal}{{Theoretical Computer Science}}}
  \bibinfo{volume}{240}(\bibinfo{number}{1}), pp. \bibinfo{pages}{177--213},
  \doi{10.1016/S0304-3975(99)00231-5}.

\bibitemdeclare{article}{vanglabbeek01}
\bibitem{vanglabbeek01}
\bibinfo{author}{R.~J. \surnamestart van Glabbeek\surnameend}
  (\bibinfo{year}{2001}): \emph{\bibinfo{title}{{The Linear Time~--~Branching
  Time Spectrum~I: The Semantics of Concrete, Sequential Processes}}}.
\newblock {\sl \bibinfo{journal}{Handbook of Process Algebra}}, pp.
  \bibinfo{pages}{3--99}, \doi{10.1016/B978-044482830-9/50019-9}.

\bibitemdeclare{inproceedings}{vG94a}
\bibitem{vG94a}
\bibinfo{author}{R.~J.~van \surnamestart Glabbeek\surnameend}
  (\bibinfo{year}{1994}): \emph{\bibinfo{title}{On the expressiveness of {ACP}
  (extended abstract)}}.
\newblock In: {\sl \bibinfo{booktitle}{Proceedings of ACP}},
  \bibinfo{series}{Workshops in Computing}, \bibinfo{publisher}{Springer}, pp.
  \bibinfo{pages}{188--217}, \doi{10.1007/978-1-4471-2120-6\_8}.

\bibitemdeclare{inproceedings}{vG12}
\bibitem{vG12}
\bibinfo{author}{R.J.~van \surnamestart Glabbeek\surnameend}
  (\bibinfo{year}{2012}): \emph{\bibinfo{title}{{Musings on Encodings and
  Expressiveness}}}.
\newblock In: {\sl \bibinfo{booktitle}{Proceedings of EXPRESS/SOS}}, {\sl
  \bibinfo{series}{EPTCS}}~\bibinfo{volume}{89}, pp. \bibinfo{pages}{81--98},
  \doi{10.4204/EPTCS.89.7}.

\bibitemdeclare{article}{Gorla10a}
\bibitem{Gorla10a}
\bibinfo{author}{D.~\surnamestart Gorla\surnameend} (\bibinfo{year}{2010}):
  \emph{\bibinfo{title}{Towards a unified approach to encodability and
  separation results for process calculi}}.
\newblock {\sl \bibinfo{journal}{Information and Computation}}
  \bibinfo{volume}{208}(\bibinfo{number}{9}), pp. \bibinfo{pages}{1031--1053},
  \doi{10.1016/j.ic.2010.05.002}.

\bibitemdeclare{article}{gorla2014abstraction}
\bibitem{gorla2014abstraction}
\bibinfo{author}{D.~\surnamestart Gorla\surnameend} \&
  \bibinfo{author}{U.~\surnamestart Nestmann\surnameend}
  (\bibinfo{year}{2014}): \emph{\bibinfo{title}{Full abstraction for
  expressiveness: history, myths and facts}}.
\newblock {\sl \bibinfo{journal}{Mathematical Structures in Computer Science}},
  pp. \bibinfo{pages}{1--16}, \doi{10.1017/S0960129514000279}.

\bibitemdeclare{inproceedings}{cspToCcs15}
\bibitem{cspToCcs15}
\bibinfo{author}{M.~\surnamestart Hatzel\surnameend},
  \bibinfo{author}{C.~\surnamestart Wagner\surnameend},
  \bibinfo{author}{K.~\surnamestart Peters\surnameend} \&
  \bibinfo{author}{U.~\surnamestart Nestmann\surnameend}
  (\bibinfo{year}{2015}): \emph{\bibinfo{title}{{Encoding CSP into CCS}}}.
\newblock In: {\sl \bibinfo{booktitle}{Proceedings of EXPRESS/SOS}},
  \bibinfo{series}{\thisvolume[EPTCS]{7}}.

\bibitemdeclare{article}{milnerParrowWalker92}
\bibitem{milnerParrowWalker92}
\bibinfo{author}{R.~\surnamestart Milner\surnameend},
  \bibinfo{author}{J.~\surnamestart Parrow\surnameend} \&
  \bibinfo{author}{D.~\surnamestart Walker\surnameend} (\bibinfo{year}{1992}):
  \emph{\bibinfo{title}{{A Calculus of Mobile Processes, Part I and II}}}.
\newblock {\sl \bibinfo{journal}{Information and Computation}}
  \bibinfo{volume}{100}(\bibinfo{number}{1}), pp. \bibinfo{pages}{1--77},
  \doi{10.1016/0890-5401(92)90008-4}.

\bibitemdeclare{article}{mitchell93}
\bibitem{mitchell93}
\bibinfo{author}{J.~C. \surnamestart Mitchell\surnameend}
  (\bibinfo{year}{1993}): \emph{\bibinfo{title}{{On abstraction and the
  expressive power of programming languages}}}.
\newblock {\sl \bibinfo{journal}{Science of Computer Programming}}
  \bibinfo{volume}{21}(\bibinfo{number}{2}), pp. \bibinfo{pages}{141--163},
  \doi{10.1016/0167-6423(93)90004-9}.

\bibitemdeclare{article}{Nestmann00}
\bibitem{Nestmann00}
\bibinfo{author}{U.~\surnamestart Nestmann\surnameend} (\bibinfo{year}{2000}):
  \emph{\bibinfo{title}{{What is a ``Good'' Encoding of Guarded Choice?}}}
\newblock {\sl \bibinfo{journal}{Information and Computation}}
  \bibinfo{volume}{156}(\bibinfo{number}{1-2}), pp. \bibinfo{pages}{287--319},
  \doi{10.1006/inco.1999.2822}.

\bibitemdeclare{inproceedings}{Nestmann06}
\bibitem{Nestmann06}
\bibinfo{author}{U.~\surnamestart Nestmann\surnameend} (\bibinfo{year}{2006}):
  \emph{\bibinfo{title}{{Welcome to the Jungle: A Subjective Guide to Mobile
  Process Calculi}}}.
\newblock In: {\sl \bibinfo{booktitle}{Proceedings of CONCUR}}, {\sl
  \bibinfo{series}{\rm LNCS}} \bibinfo{volume}{4137},
  \bibinfo{publisher}{Springer}, pp. \bibinfo{pages}{52--63},
  \doi{10.1007/11817949\_4}.

\bibitemdeclare{article}{nestmannPierce00}
\bibitem{nestmannPierce00}
\bibinfo{author}{U.~\surnamestart Nestmann\surnameend} \&
  \bibinfo{author}{B.~C. \surnamestart Pierce\surnameend}
  (\bibinfo{year}{2000}): \emph{\bibinfo{title}{{Decoding Choice Encodings}}}.
\newblock {\sl \bibinfo{journal}{Information and Computation}}
  \bibinfo{volume}{163}(\bibinfo{number}{1}), pp. \bibinfo{pages}{1--59},
  \doi{10.1006/inco.2000.2868}.

\bibitemdeclare{book}{isabelle02}
\bibitem{isabelle02}
\bibinfo{author}{T.~\surnamestart Nipkow\surnameend}, \bibinfo{author}{L.~C.
  \surnamestart Paulson\surnameend} \& \bibinfo{author}{M.~\surnamestart
  Wenzel\surnameend} (\bibinfo{year}{2002}):
  \emph{\bibinfo{title}{{Isabelle/HOL: a proof assistant for higher-order
  logic}}}.
\newblock \bibinfo{volume}{2283}, \bibinfo{publisher}{Springer Science \&
  Business Media}, \doi{10.1007/3-540-45949-9\_1}.

\bibitemdeclare{article}{Palamidessi03}
\bibitem{Palamidessi03}
\bibinfo{author}{C.~\surnamestart Palamidessi\surnameend}
  (\bibinfo{year}{2003}): \emph{\bibinfo{title}{{Comparing The Expressive Power
  Of The Synchronous And Asynchronous Pi-Calculi}}}.
\newblock {\sl \bibinfo{journal}{Mathematical Structures in Computer Science}}
  \bibinfo{volume}{13}(\bibinfo{number}{5}), pp. \bibinfo{pages}{685--719},
  \doi{10.1017/S0960129503004043}.

\bibitemdeclare{inproceedings}{PalamidessiSVV06}
\bibitem{PalamidessiSVV06}
\bibinfo{author}{C.~\surnamestart Palamidessi\surnameend},
  \bibinfo{author}{V.~A. \surnamestart Saraswat\surnameend},
  \bibinfo{author}{F.~D. \surnamestart Valencia\surnameend} \&
  \bibinfo{author}{B.~\surnamestart Victor\surnameend} (\bibinfo{year}{2006}):
  \emph{\bibinfo{title}{{On the Expressiveness of Linearity vs Persistence in
  the Asychronous Pi-Calculus}}}.
\newblock In: {\sl \bibinfo{booktitle}{Proceedings of LICS}},
  \bibinfo{publisher}{IEEE Computer Society}, pp. \bibinfo{pages}{59--68},
  \doi{10.1109/LICS.2006.39}.

\bibitemdeclare{article}{Parrow08}
\bibitem{Parrow08}
\bibinfo{author}{J.~\surnamestart Parrow\surnameend} (\bibinfo{year}{2008}):
  \emph{\bibinfo{title}{{Expressiveness of Process Algebras}}}.
\newblock {\sl \bibinfo{journal}{Electronic Notes Theoretical Computer
  Science}} \bibinfo{volume}{209}, pp. \bibinfo{pages}{173--186},
  \doi{10.1016/j.entcs.2008.04.011}.

\bibitemdeclare{article}{parrow2014abstraction}
\bibitem{parrow2014abstraction}
\bibinfo{author}{J.~\surnamestart Parrow\surnameend} (\bibinfo{year}{2014}):
  \emph{\bibinfo{title}{General conditions for full abstraction}}.
\newblock {\sl \bibinfo{journal}{Mathematical Structures in Computer Science}},
  pp. \bibinfo{pages}{1--3}, \doi{10.1017/S0960129514000280}.

\bibitemdeclare{inproceedings}{parrowCoupled92}
\bibitem{parrowCoupled92}
\bibinfo{author}{J.~\surnamestart Parrow\surnameend} \&
  \bibinfo{author}{P.~\surnamestart Sj\"odin\surnameend}
  (\bibinfo{year}{1992}): \emph{\bibinfo{title}{Multiway synchronization
  verified with coupled simulation}}.
\newblock In: {\sl \bibinfo{booktitle}{Proceedings of CONCUR}}, {\sl
  \bibinfo{series}{LNCS}} \bibinfo{volume}{630},
  \bibinfo{organization}{Springer}, pp. \bibinfo{pages}{518--533},
  \doi{10.1007/BFb0084813}.

\bibitemdeclare{phdthesis}{perez09}
\bibitem{perez09}
\bibinfo{author}{J.~A. \surnamestart Perez\surnameend} (\bibinfo{year}{2009}):
  \emph{\bibinfo{title}{{Higher-Order Concurrency: Expressiveness and
  Decidability Results}}}.
\newblock \bibinfo{type}{Ph.d. thesis}, \bibinfo{school}{University of
  Bologna}.

\bibitemdeclare{phdthesis}{peters12}
\bibitem{peters12}
\bibinfo{author}{K.~\surnamestart Peters\surnameend} (\bibinfo{year}{2012}):
  \emph{\bibinfo{title}{{Translational Expressiveness}}}.
\newblock Ph.D. thesis, \bibinfo{school}{TU Berlin}.
\newblock \urlprefix\url{http://opus.kobv.de/tuberlin/volltexte/2012/3749/}.

\bibitemdeclare{inproceedings}{PN12}
\bibitem{PN12}
\bibinfo{author}{K.~\surnamestart Peters\surnameend} \&
  \bibinfo{author}{U.~\surnamestart Nestmann\surnameend}
  (\bibinfo{year}{2012}): \emph{\bibinfo{title}{{Is it a ``Good'' Encoding of
  Mixed Choice?}}}
\newblock In: {\sl \bibinfo{booktitle}{Proceedings of FOSSACS}}, {\sl
  \bibinfo{series}{\rm LNCS}} \bibinfo{volume}{7213},
  \bibinfo{publisher}{Springer}, pp. \bibinfo{pages}{210--224},
  \doi{10.1007/978-3-642-28729-9\_14}.

\bibitemdeclare{inproceedings}{petersNestmannGoltz13}
\bibitem{petersNestmannGoltz13}
\bibinfo{author}{K.~\surnamestart Peters\surnameend},
  \bibinfo{author}{U.~\surnamestart Nestmann\surnameend} \&
  \bibinfo{author}{U.~\surnamestart Goltz\surnameend} (\bibinfo{year}{2013}):
  \emph{\bibinfo{title}{{On Distributability in Process Calculi}}}.
\newblock In: {\sl \bibinfo{booktitle}{Proceedings of ESOP}}, {\sl
  \bibinfo{series}{LNCS}} \bibinfo{volume}{7792},
  \bibinfo{publisher}{Springer}, pp. \bibinfo{pages}{310--329},
  \doi{10.1007/978-3-642-37036-6\_18}.

\bibitemdeclare{inproceedings}{riecke91}
\bibitem{riecke91}
\bibinfo{author}{J.~G. \surnamestart Riecke\surnameend} (\bibinfo{year}{1991}):
  \emph{\bibinfo{title}{{Fully abstract translations between functional
  languages}}}.
\newblock In: {\sl \bibinfo{booktitle}{Proceedings of POPL}},
  \bibinfo{organization}{ACM}, pp. \bibinfo{pages}{245--254},
  \doi{10.1145/99583.99617}.

\end{thebibliography}

\end{document}